\documentclass[structabstract]{aa}  
\usepackage{graphicx}
\usepackage{natbib}
\bibpunct{(}{)}{;}{a}{}{,}
\usepackage{txfonts}
\begin{document}
   \title{Stellar substructures in the solar neighbourhood}

   \subtitle{I. Kinematic group 3 in the Geneva-Copenhagen survey}

   \author{E. Stonkut\.{e}\inst{1},
          G. Tautvai\v{s}ien\.{e}\inst{1},
          B. Nordstr\"{o}m\inst{2},
          \and
          R. \v{Z}enovien\.{e}\inst{1}
          }
   \institute{Institute of Theoretical Physics and Astronomy (ITPA), Vilnius University,
              A. Gostauto 12, LT-01108 Vilnius, Lithuania\\
              \email{edita.stonkute@tfai.vu.lt, grazina.tautvaisiene@tfai.vu.lt, 
                     renata.zenoviene@tfai.vu.lt}
         \and
        Niels Bohr Institute, Copenhagen University, Juliane Maries Vej 30, DK-2100, Copenhagen, Denmark\\
             \email{birgitta@nbi.ku.dk}
             }

   \date{Received December 29, 2011; accepted March 8, 2012}

 
  \abstract
   {Galactic Archeology is a powerful tool for investigating the formation and evolution of the Milky Way. We use this 
technique to study kinematic groups of F- and G-stars in the solar neighbourhood. From correlations between orbital 
parameters, three new coherent groups of stars were recently identified and suggested to correspond to 
remnants of disrupted satellites.}
   {We determine detailed elemental abundances in stars belonging to one of these groups and compare their chemical composition 
with Galactic disc stars. The aim is to look for possible chemical signatures that might give information about the history of this kinematic group of stars.}
   {High-resolution spectra were obtained with the FIES spectrograph at the Nordic Optical Telescope, La Palma, and analysed
with a differential model atmosphere method. Comparison stars were observed and analysed with the same method.}
   {The average value of [Fe/H] for the 20 stars investigated in this study is $-0.69\pm 0.05$~dex. 
Elemental abundances of oxygen and $\alpha$-elements are overabundant in comparison with Galactic thin-disc dwarfs and 
thin-disc chemical evolution models. This abundance pattern has similar characteristics as the 
    Galactic thick-disc.}
{The homogeneous chemical composition together with the kinematic properties and ages of stars in the 
investigated Group~3 of the Geneva-Copenhagen survey provides evidence of their common origin and possible relation 
to an ancient merging event. 
The similar chemical composition of stars in the investigated group and the thick-disc stars might suggest that their 
formation histories are linked.}


   \maketitle

\section{Introduction}

The history of our home Galaxy is complex and not fully understood. Observations and theoretical 
simulations have made much progress and provided us with tools to search for past accretion events 
in the Milky Way and beyond. The well-known current events are the Sagittarius \citep{ibata94}, 
Canis Major \citep{martin04} and Segue~2 \citep{belokurov09} dwarf spheroidal galaxies, 
merging into the Galactic disc at 
various distances. The Monoceros stream \citep{yanny03, ibata03} and  
the Orphan stream \citep{belokurov06} according to some studies are interpreted as tidal debris from the Canis Major and 
Ursa Major~II dwarf galaxies, respectively (see \citealt{penarubia05, fellhauer07}; and the review of \citealt{helmi08}).
Accreted substructures are found  also in other galaxies, such as the Andromeda galaxy \citep{ibata01, mcconnachie09}, 
NGC~5907 \citep{martinez08}, and NGC~ 4013 \citep{martinez09}. 

\citet{helmi06} have used a homogeneous data set of about
13.240 F- and G-type stars from the \citet{nordstrom04} catalogue, which has complete kinematic,
metallicity, and age parameters, to search for signatures of past accretions in the Milky Way. From correlations 
between orbital parameters, such as apocentre (A), pericentre (P), and \textit{z}-angular momentum ($L_z$), 
the so- called APL space, Helmi et al.\ identified three new coherent groups of stars and suggested that those 
might correspond to remains of disrupted satellites. In the \textit{U--V} plane, the investigated stars are 
distributed in a banana-shape, whereas the disc stars define a centrally concentrated clump (Fig.~1). At the same time, 
in the \textit{U--W} plane the investigated stars populate mostly the outskirts of the distributions. Both the \textit{U} 
and \textit{W} distributions are very symmetric. The investigated stars have a lower mean rotational velocity in 
comparison to the Milky Way disc stars, as we can see in the \textit{W--V} plane. These characteristics are typical for stars 
associated with accreted satellite galaxies \citep{helmi08, villalobos09}. 

   \begin{figure*}
   \centering
   \includegraphics[width=\textwidth]{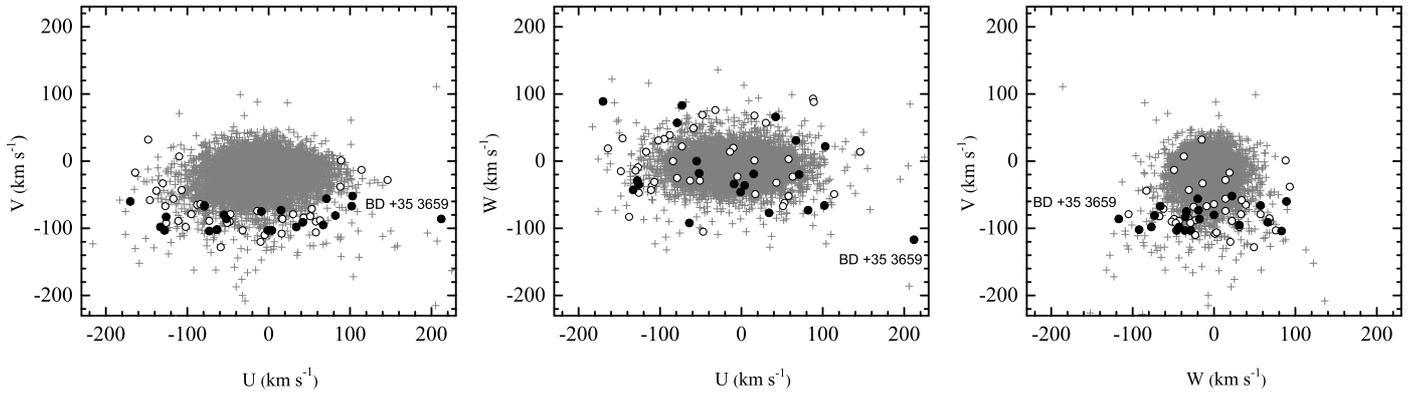}
      \caption{Velocity distribution for all stars in the \citet{holmberg09} sample 
(plus signs), stars of Group~3 (circles) and the investigated stars (filled circles). 
              }
         \label{Fig.1}
   \end{figure*}

   \begin{figure*}
   \centering
 \includegraphics[width=0.95\textwidth]{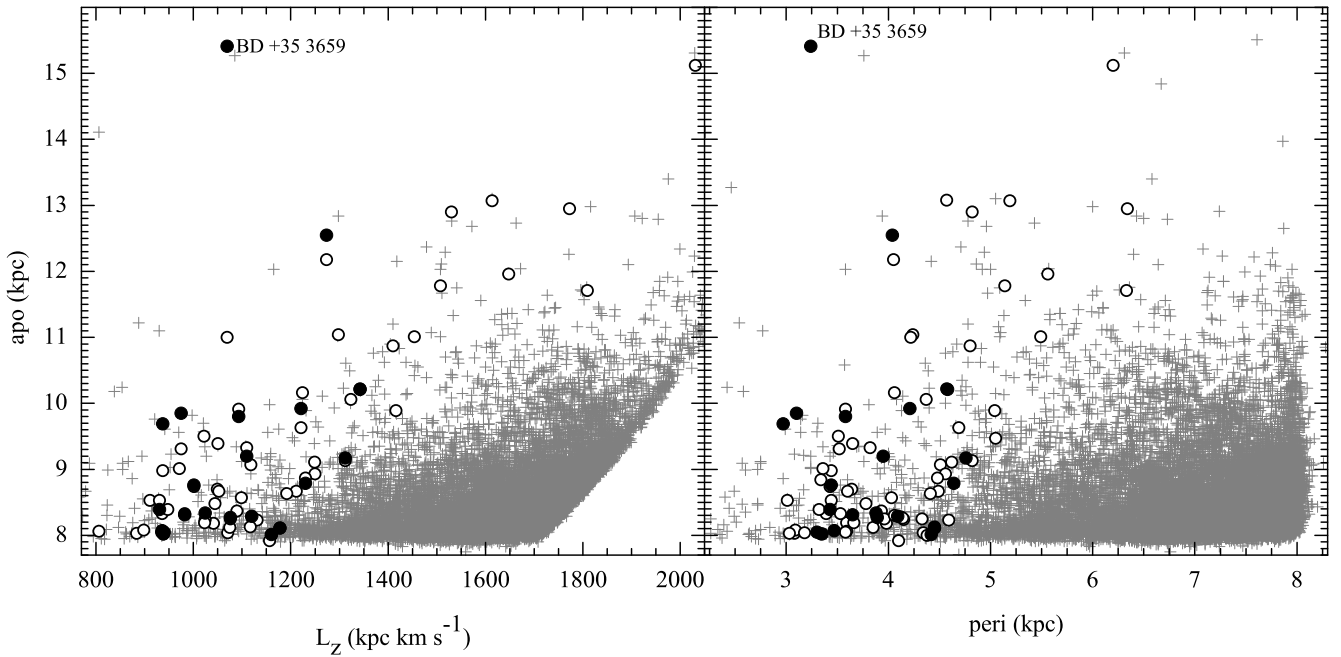}

      \caption{Distribution for the stars in the APL space. Plus signs denote the \citet{holmberg09}
      sample, circles -- Group~3, filled circles -- investigated stars. Note that the investigated stars as well as all Group 3 stars are distributed
      in APL space with constant eccentricity. 
              }
         \label{Fig.2}
   \end{figure*}

Stars in the identified groups cluster not only around regions of roughly constant eccentricity 
($0.3 \le \epsilon < 0.5$) and  
have distinct kinematics, but have also distinct metallicities [Fe/H] and age distributions.
One of the parameters according to which the stars were divided into three groups was metallicity.
Group~3, which we investigate in this work, is the most metal-deficient and consists of 68 stars. According to the 
\citet{nordstrom04} catalogue, its mean photometric metallicity, [Fe/H], is about $-0.8$~dex and the age is 
about 14~Gyr. Group 3 also differs from the other two groups by slightly different kinematics, particularly 
in the vertical ($z$) direction. \citet{holmberg09} updated and improved the parameters for the stars in the 
\citet{nordstrom04} catalogue and we use those values throughout. 

In Fig.~1 we show the Galactic disc stars from \citet{holmberg09}. Stars belonging to Group~3 in \citeauthor{helmi06} 
are marked with open and filled circles (the latter are used to mark stars investigated in our work). Evidently, stars belonging to Group 3 
have a different distribution in the velocity space in comparison to other stars of the Galactic disc. 
In Fig.~2, the stars are shown in the APL space. 
 
From high-resolution spectra we have measured abundances of iron group and $\alpha$-elements 
in 21 stars belonging to Group~3 to check the homogeneity of their chemical composition and compare them with Galactic disc stars. 
The $\alpha$-element-to-iron ratios are very sensitive indicators of galactic evolution (\citealt{pagel95, fuhrmann98, reddy06, tautvaisiene07, tolstoy09}
and references therein). If stars have been formed in different 
environments they normally have different $\alpha$-element-to-iron ratios for a given metallicity. 

 
\begin{table*}
	\centering
\begin{minipage}{150mm}
\caption{Parameters of the programme and comparison stars.}
\label{table:1} 
\begin{tabular}{lcrrrrrrccrrrrr}
\hline
\hline
Star &  Sp. type &	Age & $M_{\rm V}$ & d & $U_{\rm LSR}$ &  $V_{\rm LSR}$ &  $W_{\rm LSR}$ & $e$ & $z_{\rm max}$ & $R_{\min}$ & $R_{\max}$  \\
   			&	 &  Gyr & mag           & pc & km s$^{-1}$   &  km s$^{-1}$ & km s$^{-1}$  &        & kpc & kpc & kpc \\
 \hline
 \noalign{\smallskip}
 \object{HD 967}      & G5       & 9.9	& 5.23 & 43  & --55  & --80  & 0 & 0.34 & 0.12 &	4.09 &	8.29 \\
 \object{HD 17820}    & G5		& 11.2	& 4.45 & 61  & 34	 & --98	 &--77  & 0.39 & 1.62 &	3.65 &	8.31 \\ 
 \object{HD 107582}   & G2V      & 9.4	& 5.18 & 41  &--1    &--103  &--46  & 0.41 & 0.76 &	3.35 &	8.02 \\
 \object{BD +73 566}  & G0       & ...   & 5.14 & 67  &--52   &--86   &--18  & 0.36 & 0.18 & 3.89 &	8.27 \\
 \object{BD +19 2646} & G0       & ...   & 5.53 & 74  & 103   &--52   & 22   & 0.38 & 0.64 &	4.58 &	10.21 \\
 \object{HD 114762}   & F9V      & 10.6	& 4.36 & 39  & --79  & --66  & 57   & 0.31 & 1.57 &	4.64 &	8.79 \\  
 \object{HD 117858}   & G0       & 11.7	& 4.02 & 61  & 71    &--56   &--20  & 0.32 & 0.21 & 4.76 &	9.17 \\
 \object{BD +13 2698} & F9V      & 14.2	& 4.52 & 93  & 102   & --67  & --66 & 0.40 & 1.60 &	4.21 &	9.92 \\
 \object{BD +77 0521} & G5       & 14.5	& 5.27 & 68  & 4     & --103 & --36 & 0.42 & 0.48 &	3.30 &	8.05 \\
 \object{HD 126512}   & F9V      & 11.1	& 4.01 & 45  & 82    & --81  & --73 & 0.40 & 1.69 &	3.95 &	9.20 \\
 \object{HD 131597}   & G0       & ...	& 3.06 & 119 & --133 & --98  & --43 & 0.52 & 0.81 &	3.10 &	9.85 \\
 \object{BD +67 925}  & F8       & 13    & 4.14 & 139 & --128 & --103 & --29 & 0.53 & 0.43 &	2.97 &	9.69 \\
 \object{HD 159482}   & G0V      & 10.9  & 4.82 & 52  & --170 & --60  & 89   & 0.51 & 3.67 &	4.04 &	12.55 \\
 \object{HD 170737}   & G8III-IV & ...   & 2.88 & 112 & --64  & --102 & --92 & 0.40 & 2.61 &	3.47 &	8.07 \\
 \object{BD +35 3659} & F1       & 0.9	& 5.32 & 96  & 212   & --86  & --117& 0.65 & 5.50 &	3.24 &	15.41 \\
 \object{HD 201889}   & G1V      & 14.5	& 4.40 & 54  & --126 & --83  & --35 & 0.46 & 0.56 & 3.58 &  9.80 \\
 \object{HD 204521}   & G5       & 2.1   & 5.18 & 26  & 15    & --73  & --19 & 0.29 & 0.18 & 4.45 &	8.11 \\
 \object{HD 204848}   & G0       & ... 	& 1.98 & 122 & 42    & --91  & 66   & 0.36 & 1.77 &	3.88 &	8.34 \\
 \object{HD 212029}   & G0       & 13.1	& 4.66 & 59  & 67    & --95  & 31   & 0.44 & 0.77 &	3.44 &	8.76 \\
 \object{HD 222794}   & G2V      & 12.1  & 3.83 & 46  & --73  & --104 & 83   & 0.42 & 3.02 &	3.43 &	8.39 \\
 \object{HD 224930}   & G5V		& 14.7	& 5.32 & 12  & --9   &--75   & --34 & 0.29 & 0.44 & 4.42 &	8.01 \\
 \noalign{\smallskip}
 \hline
 \noalign{\smallskip}
 \object{HD 17548}    & F8       & 6.9	& 4.46 & 55  & --14  & 31	 & 32	& 0.15 & 0.88 &	8.02 &	10.85 \\
 \object{HD 150177}   & F3V		& 5.7	& 3.33 & 40  & --7	 & --23	 & --24 & 0.07 & 0.25 &	6.95 &	7.97 \\ 
 \object{HD 159307}   & F8       & 5.2	& 3.33 & 65  & --14  & --21  & 0    & 0.06 & 0.11 &	7.00 &	7.95 \\
 \object{HD 165908}   & F7V      & 6.8	& 4.09 & 16  & --6   & 1     & 10   & 0.03 & 0.26 & 7.97 &  8.45 \\
 \object{HD 174912}   & F8       & 6.1   & 4.73 & 31  & --22  & 8     & --43 & 0.07 & 0.69 & 7.90 &	9.10 \\
 \object{HD 207978}   & F6IV-V   & 3.3	& 3.33 & 27  & 13    & 16    & --7  & 0.11 & 0.00 &	7.81 &	9.68 \\

\hline
\end{tabular}
\end{minipage}
\end{table*}


\section{Observations and method of analysis}

Spectra of high-resolving power ($R\approx$68\,000) in the wavelength range of 3680--7270~{\AA} were obtained 
at the Nordic Optical Telescope with the FIES  spectrograph during July 2008. Twenty-one programme and six comparison stars    
(thin-disc dwarfs) were observed.  A list of the observed stars and some of their parameters (taken from the 
\citealt{holmberg09} catalogue and Simbad) are presented in Table~1.

All spectra were exposed to reach a signal-to-noise ratio higher than 100. Reductions of CCD images were made 
with the FIES pipeline {\sc FIEStool}, which performs  a complete reduction: calculation of reference frame, bias and 
scattering subtraction, flat-field dividing, wavelength calibration and other procedures (http://www.not.iac.es/instruments/fies/fiestool). Several examples 
of  stellar spectra are presented in Fig.~3.

The spectra were analysed using a differential model atmosphere technique. 
The {\sc Eqwidth} and {\sc Spectrum} program packages, developed at the Uppsala Astronomical 
Observatory, were used to carry out the calculation of abundances from measured 
equivalent widths and synthetic spectra, respectively.
A set of plane-parallel, line-blanketed, constant-flux LTE model atmospheres 
\citep{gustafsson08} were taken from the {\sc MARCS} stellar model atmosphere and flux 
library (http://marcs.astro.uu.se/).

The Vienna Atomic Line Data Base (VALD, \citealt{piskunov95}) was extensively 
used in preparing input data for the calculations. Atomic oscillator 
strengths for the main spectral lines analysed in this study were taken from  
an inverse solar spectrum analysis performed in Kiev \citep{gurtovenko89}.
All lines used for calculations were carefully selected to avoid blending.  All line profiles in all spectra were 
hand-checked, requiring that the line profiles be sufficiently clean to provide reliable equivalent widths. 
The equivalent widths of the lines were measured by fitting of a Gaussian profile using the {\sc 4A} software 
package \citep{ilyin00}.

Initial values of the effective temperatures for the programme stars were taken from \citet{holmberg09} 
and then carefully checked and corrected if needed by 
forcing Fe~{\sc i} lines to yield no dependency of iron abundance on excitation 
potential by changing the model effective temperature. For four stars our 
effective temperature is $+100$ to $+200$~K higher than in the catalogue.  
We used the ionization equilibrium method to find surface gravities of the programme stars 
by forcing neutral and ionized iron lines to yield the same iron abundances.
Microturbulence velocity values corresponding to the minimal line-to-line  
Fe~{\sc i} abundance scattering were chosen as correct values. 

Using the $gf$ values and solar equivalent widths of analysed lines from \citet{gurtovenko89}
we obtained the solar abundances, used later for the 
differential determination of abundances in the programme stars. We used the 
solar model atmosphere from the set calculated in Uppsala with a microturbulent 
velocity of 0.8~$\rm {km~s}^{-1}$, as derived from Fe~{\sc i} lines. 

\input epsf
\begin{figure}
\epsfxsize=\hsize 
\epsfbox[10 20 280 220]{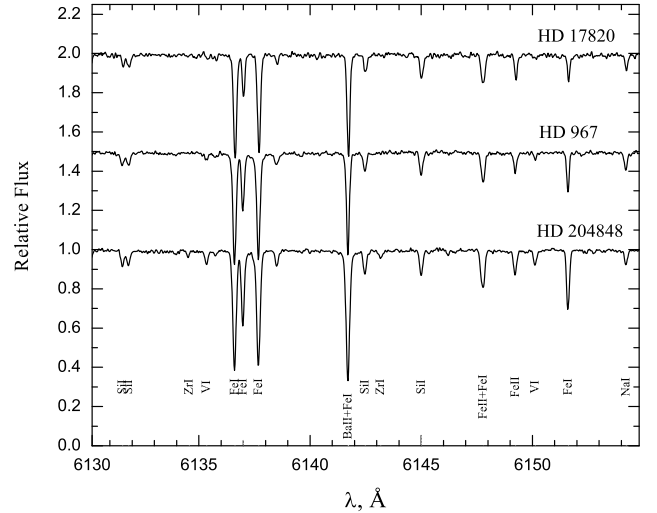} 
\caption{Samples of stellar spectra of several programme stars. An offset of 0.5 in relative flux is applied for clarity.} 
\label{fig3}
\end{figure}

In addition to thermal and microturbulent Doppler broadening of lines, atomic 
line broadening by radiation damping and van der Waals damping were considered 
in the calculation of abundances. Radiation damping parameters of lines were taken from the VALD database. 
In most cases the hydrogen pressure damping of metal lines was treated using 
the modern quantum mechanical calculations by \citet{anstee95}, 
\citet{barklem97}, and \citet{barklem98}. 
When using the \citet{unsold55} approximation, correction factors to the classical 
van der Waals damping approximation by widths 
$(\Gamma_6)$ were taken from \citet{simmons82}. For all other species a correction factor 
of 2.5 was applied to the classical $\Gamma_6$ $(\Delta {\rm log}C_{6}=+1.0$), 
following \citet{mackle75}. For lines stronger than $W=100$~m{\AA} the correction factors were selected individually by  
inspection of the solar spectrum.

The oxygen abundance was determined from the forbidden [O\,{\sc i}] line at 6300.31~\AA\ (Fig.~4). The oscillator strength 
values for \textsuperscript{58}Ni and \textsuperscript{60}Ni, which blend the oxygen line, were taken from \citet{johansson03}. 
The [O\,{\sc i}] log~$gf = -9.917$ value was calibrated by fitting to the solar spectrum \citep{kurucz05} with log~$A_{\odot}=8.83$ taken 
from \citet{grevesse00}. Stellar rotation was taken into account if needed with $v {\rm sin} i$ values from \citet{holmberg07}.
Abundances of oxygen was not determined for every star due to blending by telluric lines or weakness of the oxygen line profile. 
     
\input epsf
\begin{figure}
\epsfxsize=\hsize 
\epsfbox[15 10 290 245]{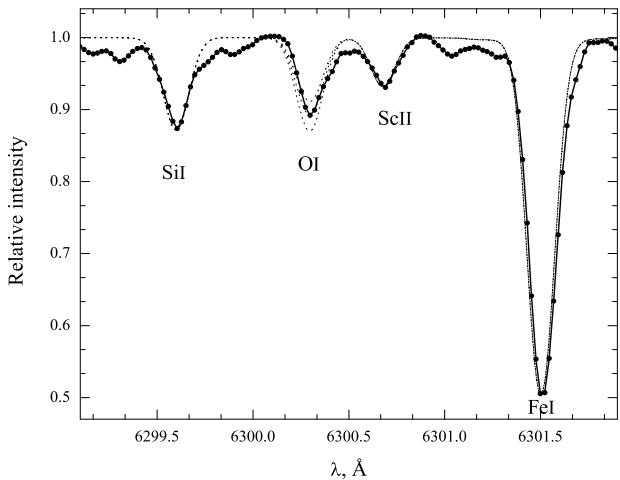} 
\caption{Fit to the forbidden [O\,{\sc i}] line at 6300.3 {\AA} in the programme star HD 204848. 
The observed spectrum is shown as a solid line with black dots. The synthetic spectra with ${\rm [O/Fe]}=0.52 \pm 0.1$ 
are shown as dashed lines.} 
\label{Fig.4}
\end{figure}

Abundances of other chemical elements were determined using equivalent widths of their lines. 
Abundances of Na and Mg were determined with non-local thermodynamical equilibrium (NLTE) taken into 
account, as described by \citet{gratton99}. The calculated corrections did not exceed $0.04$~dex for 
Na \,{\sc i} and $0.06$~dex for Mg \,{\sc i} lines. 
Abundances of sodium were determined from equivalent widths of the Na\,{\sc i} lines 
at 5148.8, 5682.6, 6154.2, and 6160.8~{\AA}; magnesium from the Mg\,{\sc i} lines at 4730.0, 5711.1, 
6318.7, and 6319.2~{\AA}; and that of aluminum from the Al\,{\sc i} lines at 6696.0, 6698.6, 7084.6, and 7362.2~{\AA}.

\subsection{Estimation of uncertainties}

   \begin{table}
   \centering
   \begin{minipage}{80mm}
      \caption{Effects on derived abundances resulting from model changes for the star HD~224930.} 
        \label{table:2}
      \[
         \begin{tabular}{lrrc}
            \hline
	    \hline
            \noalign{\smallskip}
Ion & ${ \Delta T_{\rm eff} }\atop{ -100 {\rm~K} }$ & 
            ${ \Delta \log g }\atop{ -0.3 }$ & 
            ${ \Delta v_{\rm t} }\atop{ -0.3~{\rm km~s}^{-1}}$ \\ 
            \noalign{\smallskip}
            \hline
            \noalign{\smallskip}
[O\,{\sc i}] 	 &	  0.00	&	--0.10&	--0.01\\
Na\,{\sc i} 	&	--0.06	&	0.01	&	0.00	\\
Mg\,{\sc i} 	&	--0.04	&	0.01	&	0.01	\\
Al\,{\sc i} 	&	--0.05	&	0.00	&	0.00	\\
Si\,{\sc i} 	&	--0.01	&	--0.03&	0.01	\\
Ca\,{\sc i} 	&	--0.07	&	0.02	&	0.04	\\
Sc\,{\sc ii}	&	--0.01 	& 	--0.13&	0.02	\\
Ti\,{\sc i} 	&	--0.10	&	0.01	&	0.03	\\
Ti\,{\sc ii}	&	--0.01	&	--0.13&	0.03	\\
V\,{\sc i}  	&	--0.12	&	0.00	&	0.00	\\
Cr\,{\sc i} 	&	--0.09	&	0.01	&	0.05	\\
Fe\,{\sc i} 	&	--0.08	&	0.00	&	0.05	\\
Fe\,{\sc ii} 	&	0.04	        &	--0.13&	0.04	\\
Co\,{\sc i} 	&	--0.07	&	--0.02&	0.01	\\
Ni\,{\sc i} 	&	--0.05	&	--0.01&	0.04	\\
            \hline
         \end{tabular}
      \]
\end{minipage}
   \end{table}
The uncertainties in abundances are due to several sources: uncertainties caused by 
analysis of individual lines, including random errors of atomic data and continuum 
placement and  uncertainties in the stellar parameters.
The sensitivity of the abundance 
estimates to changes in the atmospheric parameters by the assumed errors $\Delta$[El/H]
are illustrated  for the star HD\, 224930 (Table~2). Clearly, possible 
parameter errors do not affect the abundances seriously; the element-to-iron 
ratios, which we use in our discussion, are even less sensitive. 

The scatter of the deduced abundances from different spectral lines $\sigma$
gives an estimate of the uncertainty due to the random errors. The mean value 
of  $\sigma$ is 0.05~dex, thus the uncertainties in the derived abundances that 
are the result of random errors amount to approximately this value.

\input epsf
\begin{figure}
\epsfxsize=\hsize 
\epsfbox[15 10 300 240]{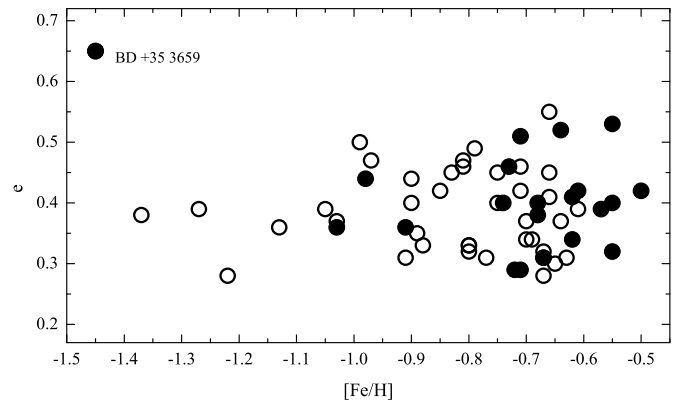} 
\caption{Diagram of orbital eccentricity \textit{e} vs. [Fe/H] for all stars of Group 3 (circles) and 
those investigated in this work (filled circles).} 
\label{Fig.5}
\end{figure}

\input epsf
\begin{figure}
\epsfxsize=\hsize 
\epsfbox[15 10 390 335]{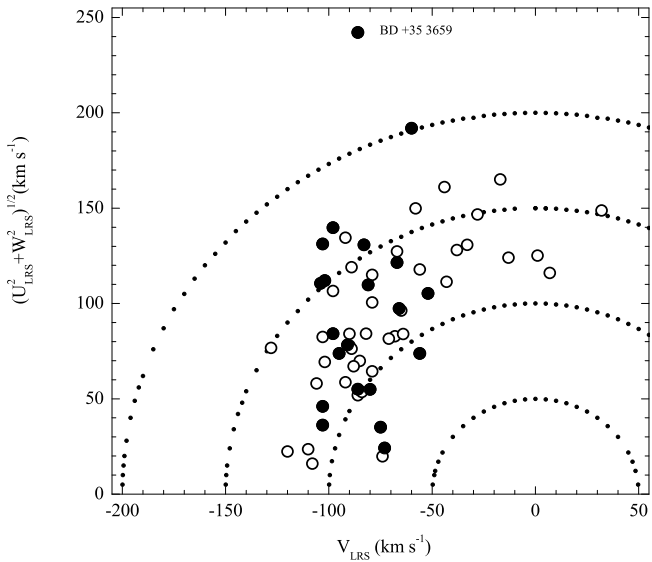} 
\caption{Toomre diagram of all stars of Group 3 (circles) and 
those investigated in this work (filled circles). Dotted lines 
indicate constant values of total space velocity in steps of 50 km s$^{-1}$.} 
\label{Fig.6}
\end{figure}

 \begin{table*}
	\centering
\begin{minipage}{190mm}
\caption{Main atmospheric parameters and elemental abundances of the programme and comparison stars.}
\label{table:3} 
\begin{tabular}{lccccccccccccccc}
\hline\hline   
Star & $T_{\rm eff}$ & log~$g$ & $v_{t}$ & [Fe/H] & $\sigma_{\rm Fe I}$ & ${\rm n}_{\rm Fe I}$ & $\sigma_{\rm Fe II}$ & ${\rm n}_{\rm Fe II}$& [O/Fe]& [Na/Fe]& $\sigma$& n& [Mg/Fe]& $\sigma$& n\\
   & K             &       & km s$^{-1}$   &      &      &    &    &  \\
 \hline
\noalign{\smallskip}
 HD 967      & 5570 & 4.3 & 0.9 & --0.62 & 0.05 & 38 & 0.04 &  7 & ...	 	& 0.04   & 0.03 & 3 & 0.33 & 0.04 & 4\\
 HD 17820    & 5900 & 4.2 & 1.0 & --0.57 & 0.05 & 29 & 0.01 &  6 & ...	 	& 0.06   & 0.04 & 3 & 0.25 & 0.06 & 3\\
 HD 107582   & 5600 & 4.2 & 1.0 & --0.62 & 0.05 & 32 & 0.07 &  5 & 0.39 	& 0.06   & 0.04 & 4 & 0.30 & 0.04 & 3\\
 BD +73 566  & 5580 & 3.9 & 0.9 & --0.91 & 0.05 & 31 & 0.02 &  6 & ...	 	& 0.14   & 0.03 & 2 & 0.43 & 0.05 & 3\\
 BD +19 2646 & 5510 & 4.1 & 0.9 & --0.68 & 0.04 & 31 & 0.04 &  5 & 0.55 	& 0.10   & 0.08 & 3 & 0.38 & 0.06 & 4\\
 HD 114762   & 5870 & 3.8 & 1.0 & --0.67 & 0.05 & 32 & 0.03 &  7 & ...		& 0.09   & 0.03 & 3 & 0.33 & 0.05 & 4\\  
 HD 117858   & 5740 & 3.8 & 1.2 & --0.55 & 0.04 & 34 & 0.03 &  6 & 0.32		& 0.08   & 0.02 & 3 & 0.29 & 0.04 & 3\\
 BD +13 2698 & 5700 & 4.0 & 1.0 & --0.74 & 0.06 & 28 & 0.05 &  5 & ...		& 0.02   & 0.02 & 2 & 0.34 & 0.05 & 4\\
 BD +77 0521 & 5500 & 4.0 & 1.1 & --0.50 & 0.07 & 24 & 0.05 &  5 & ...		& --0.02 & ...  & 1 & 0.25 & 0.04 & 4\\
 HD 126512   & 5780 & 3.9 & 1.1 & --0.55 & 0.05 & 27 & 0.03 &  6 & 0.41 	& 0.10	 & 0.02 & 3 & 0.30 & 0.07 & 3\\
 HD 131597   & 5180 & 3.5 & 1.1 & --0.64 & 0.04 & 32 & 0.03 &  6 & ...		& 0.12	 & 0.01 & 4 & 0.37 & 0.05 & 4\\
 BD +67 925  & 5720 & 3.5 & 1.2 & --0.55 & 0.05 & 24 & 0.03 &  6 & 0.37		& 0.04   & 0.02 & 2 & 0.35 & 0.06 & 3\\
 HD 159482   & 5730 & 4.1 & 1.0 & --0.71 & 0.05 & 26 & 0.01 &  5 & 0.42		& 0.13   & 0.05 & 4 & 0.31 & 0.03 & 4\\
 HD 170737   & 5100 & 3.3 & 1.0 & --0.68 & 0.04 & 29 & 0.05 &  6 & ...		& 0.11   & 0.02 & 4 & 0.30 & 0.07 & 3\\
 BD +35 3659\tablefootmark{a} & 5850 & 3.9 & 0.9 & --1.45 & 0.04 & 25 & 0.04 &  4 & ... 	& 0.04   & 0.05 & 3 & 0.30 & 0.06 & 3\\
 HD 201889   & 5700 & 3.8 & 0.9 & --0.73 & 0.05 & 30 & 0.03 &  4 & 0.58		& 0.08   & 0.04 & 3 & 0.32 & 0.03 & 4\\
 HD 204521   & 5680 & 4.3 & 1.0 & --0.72 & 0.05 & 30 & 0.05 &  5 & ...		& 0.06   & 0.02 & 3 & 0.29 & 0.04 & 3\\
 HD 204848   & 4900 & 2.3 & 1.2 & --1.03 & 0.04 & 31 & 0.05 &  7 & 0.52		& 0.01   & 0.04 & 3 & 0.43 & 0.03 & 4\\
 HD 212029   & 5830 & 4.2 & 0.9 & --0.98 & 0.02 & 20 & 0.01 &  2 & ...		& 0.10   & 0.04 & 3 & 0.37 & 0.07 & 4\\
 HD 222794   & 5560 & 3.7 & 1.1 & --0.61 & 0.04 & 30 & 0.05 &  6 & ...		& 0.09	 & 0.05 & 4 & 0.37 & 0.07 & 4\\
 HD 224930   & 5470 & 4.2 & 0.9 & --0.71 & 0.05 & 35 & 0.05 &  6 & 0.45		& 0.08   & 0.04 & 3 & 0.42 & 0.04 & 4\\
\noalign{\smallskip}
\hline
 \noalign{\smallskip}
 HD 17548    & 6030 & 4.1 & 1.0 & --0.49 & 0.05 & 32 & 0.03 & 7 & 0.16 		& --0.02 & 0.04 & 3 & 0.07 & 0.06 & 4\\
 HD 150177   & 6300 & 4.0 & 1.5 & --0.50 & 0.04 & 23 & 0.05 & 4 & ... 		& 0.07   & 0.02 & 3 & 0.18 & 0.04 & 3\\
 HD 159307   & 6400 & 4.0 & 1.6 & --0.60 & 0.04 & 17 & 0.04 & 4 & ...		& 0.12   & 0.07 & 3 & 0.28 & 0.03 & 4\\
 HD 165908   & 6050 & 3.9 & 1.1 & --0.52 & 0.04 & 24 & 0.03 & 7 & ...		& 0.02   & 0.02 & 3 & 0.20 & 0.07 & 4\\
 HD 174912   & 5860 & 4.1 & 0.8 & --0.42 & 0.04 & 33 & 0.04 & 6 & 0.10 		& 0.02   & 0.01 & 3 & 0.08 & 0.05 & 4\\
 HD 207978   & 6450 & 3.9 & 1.6 & --0.50 & 0.04 & 22 & 0.04 & 7 & ...		& 0.09   & 0.05 & 3 & 0.28 & 0.06 & 4\\
\hline
\end{tabular}
\end{minipage}
\tablefoot{\tablefoottext{a}{Probably not a member of Group~3. }}
\end{table*}

\section{Results and discussion}

The atmospheric parameters $T_{\rm eff}$, log\,$g$, $v_{t}$, [Fe/H] and abundances of 12 chemical elements relative 
to iron [El/Fe] of the programme and comparison stars are presented in Table~3. The number of lines and 
the line-to-line scatter ($\sigma$) are presented as well.

\subsection{Comparison with previous studies} 

\begin{table}
\begin{minipage}{80mm}
\caption{Group~3 comparison with previous studies.}
\label{table:4} 
\begin{tabular}{lrrrrrr}
\hline\hline   
         &\multicolumn{2}{c}{Ours--Nissen} & \multicolumn{2}{c}{Ours--Reddy} & \multicolumn{2}{c}{Ours--Ram\'{i}rez} \\
Quantity & Diff.  & $\sigma$ &  Diff.    & $\sigma$ &  Diff.    & $\sigma$\\
\hline
T$_{\rm eff} $  & 34       & 54		 & 86   	& 33   & 47      & 45 \\
log $g$         & --0.26   & 0.16    & --0.28 	& 0.15 & --0.27  & 0.14\\
${\rm [Fe/H]}$  &  0.03    & 0.04    & 0.06 	& 0.07 & 0.10    & 0.04 \\
${\rm[Na/Fe]}$  &  0.02    & 0.11    & 0.00 	& 0.04 &  ...    & ...\\
${\rm[Mg/Fe]}$  & --0.02   & 0.05    & 0.02		& 0.01 &  ...    & ...\\
${\rm[Al/Fe]}$  &  ...     & ...     &--0.01	& 0.07 &  ...    & ...\\
${\rm[Si/Fe]}$  & --0.05   & 0.01    & 0.03 	& 0.05 &  ...    & ...\\
${\rm[Ca/Fe]}$  &  0.00    & 0.01    & 0.08		& 0.05 &  ...    & ...\\
${\rm[Sc/Fe]}$  &  ...     & ...     & --0.01   & 0.05 &  ...    & ...\\
${\rm[Ti/Fe]}$  &  0.06    & 0.10    & 0.07		& 0.06 &  ...    & ...\\
${\rm[V/Fe]}$   &  ...     & ...     & --0.01   & 0.03 &  ...    & ...\\
${\rm[Cr/Fe]}$  &  0.02    & 0.06    & 0.08	    & 0.04 &  ...    & ...\\
${\rm[Co/Fe]}$  &  ...     & ...     & --0.04	& 0.03 &  ...    & ...\\
${\rm[Ni/Fe]}$  & 0.01     & 0.05    & --0.01 	& 0.04 &  ...    & ...\\
\hline
\end{tabular}
\end{minipage}
\tablefoot{Mean differences and standard deviations of the main parameters and abundance ratios [El/Fe] for
4 stars of Group~3 that are in common with \citet{nissen10}, 7 stars in common with \citet{reddy06}, and 10
stars in common with \citet{ramirez07}.}
\end{table}

\begin{table}
\begin{minipage}{80mm}
\caption{Thin-disc stars comparison with previous studies.}
\label{table:5} 
\begin{tabular}{lrrrr}
\hline\hline   
         &\multicolumn{2}{c}{Ours--Edvardsson} & \multicolumn{2}{c}{Ours--Th\'{e}venin} \\
Quantity & Diff.  & $\sigma$ &  Diff.    & $\sigma$\\
\hline
T$_{\rm eff} $  & 86       & 66		 & 87   	& 68   \\
log $g$         & --0.18   & 0.21    & --0.06 	& 0.14 \\
${\rm [Fe/H]}$  &  0.10    & 0.04    & 0.03 	& 0.03 \\
${\rm[Na/Fe]}$  & --0.08   & 0.09    &  ...    & ... \\
${\rm[Mg/Fe]}$  & --0.02   & 0.07    &  ...    & ... \\
${\rm[Al/Fe]}$  & --0.10   & 0.07  	 &  ...    & ... \\
${\rm[Si/Fe]}$  & --0.02   & 0.02    &  ...    & ... \\
${\rm[Ca/Fe]}$  &  0.05    & 0.03    &  ...    & ... \\
${\rm[Ti/Fe]}$  & --0.02   & 0.08    &  ...    & ... \\
${\rm[Ni/Fe]}$  & --0.06   & 0.06    &  ...    & ... \\
\hline
\end{tabular}
\end{minipage}
\tablefoot{Mean differences and standard deviations of the main parameters and abundance ratios
[El/Fe] for 6 thin-disc stars that are in common with \citet{edvardsson93} and 5 stars with \citet{thevenin99}.}
\end{table}

Some stars from our sample have been previously investigated by other authors. In Table~4 we present a comparison with the results by
\citet{nissen10}, \citet{reddy06}, and with \citet{ramirez07}. \citeauthor{ramirez07} determined only the main atmospheric parameters.
The thin-disc stars we have investigated in our work for a comparison have been analysed previously by
\citet{edvardsson93} and by \citet{thevenin99}.
In Table~5 we present a comparison with the results obtained by these authors. Our [El/Fe] for the stars in common agree very well with 
other studies. Slight differences in the log\,$g$ values lie within errors of uncertainties and are caused mainly by 
differences in determination methods applied. In our work we see that titanium abundances determined using Ti{\sc i} and Ti{\sc ii} 
lines agree well and confirm the log\,$g$ values determined using iron lines. 
 
Effective temperatures for all  stars investigated here are also available in
\citet{holmberg09} and \citet{casagrande11}.
\citeauthor{casagrande11} provide astrophysical parameters for the Geneva-Copenhagen survey by applying the infrared flux method for the effective temperature determination.
In comparison to \citeauthor{holmberg09}, stars in the \citeauthor{casagrande11} catalogue are 
on average 100~K hotter. For the stars investigated here, our spectroscopic temperatures are on average $40\pm 70$~K 
hotter than in \citeauthor{holmberg09} and $60\pm 80$~K cooler than in \citeauthor{casagrande11} (BD +35 3659, which has a difference of 
340~K, was excluded from the average).  
 
[Fe/H] values for all investigated stars are available in \citet{holmberg09} as well as in \citet{casagrande11}. 
A comparison between \citeauthor{holmberg09} and \citeauthor{casagrande11} shows that the latter gives [Fe/H] values that are on average by 0.1~dex more metal-rich.
For our programme stars we obtain a difference of $0.1\pm 0.1$~dex in comparison with Holmberg et al., and no systematic difference but a scatter of 0.1~dex in comparison with \citeauthor{casagrande11}

\subsection{Comparison with the thin-and thick-disc dwarfs}  
 
  \begin{figure*}
   \centering
   \includegraphics[width=0.85\textwidth]{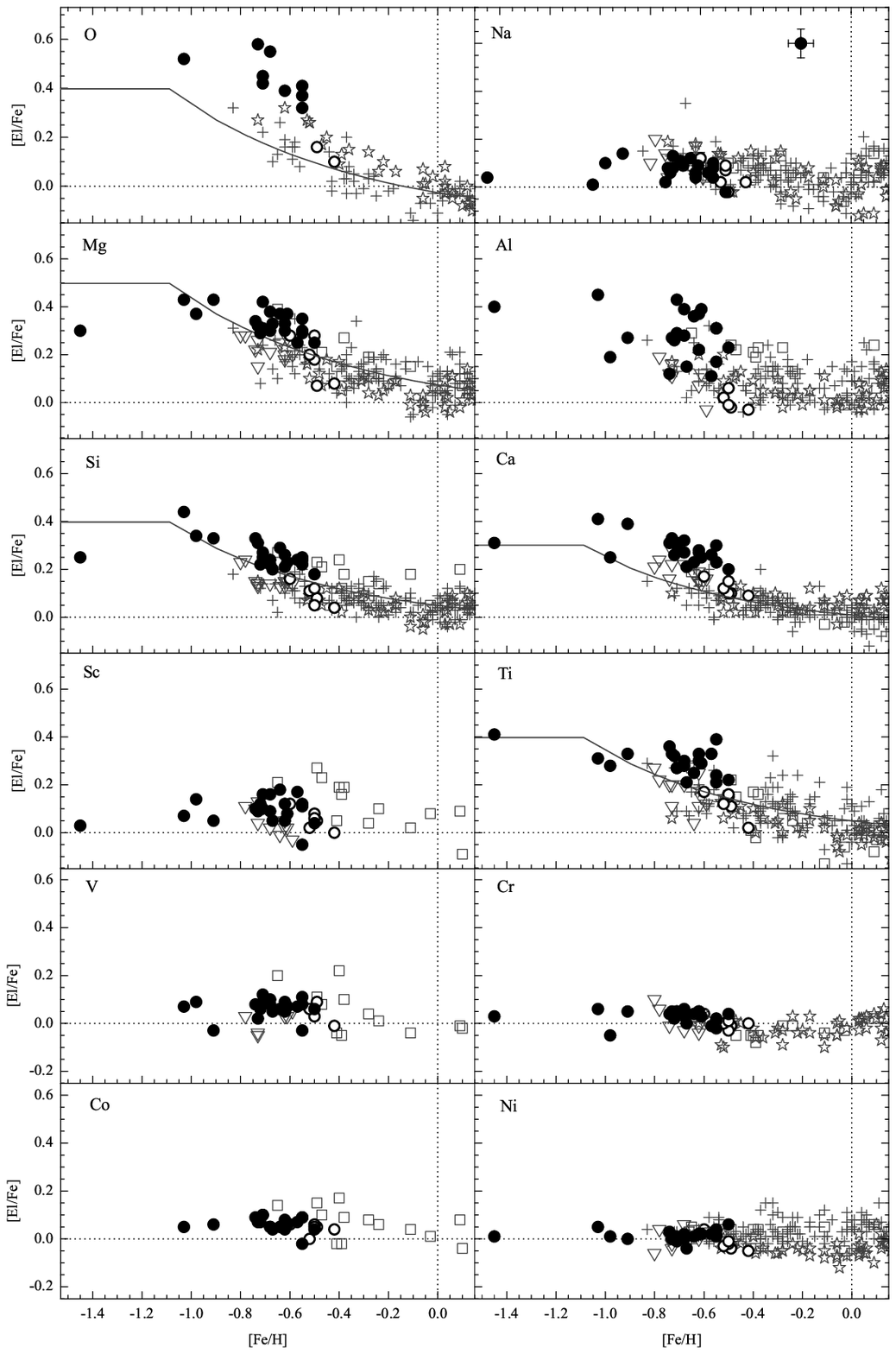}
     \caption{Comparison of elemental abundance ratios of stars in the investigated stellar group (black points) and data for Milky 
     Way thin-disc dwarfs from \citeauthor{edvardsson93} (1993, plus signs), \citeauthor{bensby05} (2005, stars), \citeauthor{reddy06} (2006, squares),
\citeauthor{zhang06} (2006, triangles), and Galactic thin disc chemical evolution models by \citeauthor{pagel95} (1995, solid lines). 
Results obtained for thin-disc dwarfs analysed in our work are shown by open circles. Average uncertainties are shown in the box for Na.}
       \label{Fig.7}
   \end{figure*}

\addtocounter{table}{-3}
\begin{table*}
	\centering
\begin{minipage}{200mm}
\caption{Continued}
\label{table:3cont} 
\begin{tabular}{lccccccccccccccc}
\hline\hline   
Star & [Al/Fe] & $\sigma$ & n & [Si/Fe] &$\sigma$ &  n& [Ca/Fe]&$\sigma$ &  n& [Sc/Fe]& $\sigma$& n& [Ti{\sc i}/Fe]& $\sigma$& n\\
 \hline
\noalign{\smallskip}
 HD 967      & 0.37 & 0.00 & 3 &	0.26 & 0.05 & 17 &	0.28 & 0.04 & 8 & 	0.12 & 0.04 & 9 &	0.33 & 0.06 & 14\\
 HD 17820    & 0.11 & 0.05 & 3 &	0.24 & 0.04 & 16 &	0.26 & 0.06 & 9 &	0.17 & 0.04 & 9 &	0.33 & 0.05 & 9\\
 HD 107582   & 0.22 & 0.05 & 3 &	0.21 & 0.05 & 16 &	0.27 & 0.06 & 5 &	0.05 & 0.05 & 8	&	0.30 & 0.05 & 7\\
 BD +73 566  & 0.27 & 0.04 & 2 &	0.33 & 0.06 & 18 &	0.39 & 0.06 & 6 &	0.05 & 0.05 & 7 &	0.33 & 0.05 & 9\\
 BD +19 2646 & 0.28 & 0.08 & 2 &	0.22 & 0.06 & 16 &	0.32 & 0.06 & 8 &	0.09 & 0.04 & 8 &	0.28 & 0.04 & 9\\
 HD 114762   & 0.15 & 0.02 & 2 &	0.20 & 0.05 & 17 &	0.21 & 0.05 & 7 &	0.05 & 0.05 & 10&	0.21 & 0.03 & 8\\  
 HD 117858   & 0.31 & 0.06 & 3 &	0.24 & 0.05 & 17 &	0.23 & 0.04 & 8 &	0.12 & 0.02 & 9 &	0.24 & 0.03 & 9\\
 BD +13 2698 & 0.12 & 0.05 & 2 &	0.33 & 0.06 & 14 &	0.31 & 0.05 & 8 &	0.10 & 0.04 & 7 &	0.36 & 0.05 & 8\\
 BD +77 0521 & 0.23 & 0.01 & 2 &	0.18 & 0.07 & 10 &	0.20 & 0.06 & 6 &	0.04 & 0.03 & 4 &	0.22 & 0.05 & 6\\
 HD 126512   & 0.17 & 0.06 & 2 &	0.25 & 0.05 & 16 &	0.23 & 0.04 & 6 &	0.11 & 0.02 & 8 &	0.21 & 0.04 & 7\\
 HD 131597   & 0.36 & 0.04 & 2 &	0.29 & 0.05 & 16 &	0.25 & 0.04 & 8 &	0.18 & 0.02 & 10&	0.25 & 0.06 & 16\\
 BD +67 925  & 0.31 & 0.00 & 2 &	0.22 & 0.08 & 17 &	0.30 & 0.06 & 8 & --0.05 & 0.05 & 4 &	0.39 & 0.03 & 3\\
 HD 159482   & 0.29 & 0.04 & 2 &	0.27 & 0.06 & 15 &	0.31 & 0.06 & 7 &	0.16 & 0.03 & 8 &	0.27 & 0.02 & 4\\
 HD 170737   & 0.39 & ...  & 1 &	0.24 & 0.04 & 15 &	0.27 & 0.06 & 7 &	0.16 & 0.02 & 8 &	0.30 & 0.06 & 12\\
 BD +35 3659 & 0.40 & 0.01 & 2 &	0.25 & 0.03 &  8 &	0.31 & 0.08 & 4 &	0.03 & 0.11 & 7 &	0.41 & 0.02 & 4\\
 HD 201889   & 0.27 & 0.01 & 3 &	0.31 & 0.05 & 15 &	0.33 & 0.08 & 6 &	0.09 & 0.03 & 9 &	0.33 & 0.05 & 9\\
 HD 204521   & 0.26 & 0.02 & 2 &	0.22 & 0.05 & 17 &	0.26 & 0.06 & 8 &	0.12 & 0.03 & 7 &	0.32 & 0.05 & 9\\
 HD 204848   & 0.45 & 0.05 & 3 &	0.44 & 0.05 & 16 &	0.41 & 0.05 & 8 &	0.07 & 0.03 & 11&	0.31 & 0.07 & 18\\
 HD 212029   & 0.19 & 0.05 & 2 &	0.34 & 0.05 & 11 &	0.25 & 0.03 & 6 &	0.14 & 0.02 & 5 &	0.28 & 0.06 & 5\\
 HD 222794   & 0.39 & 0.04 & 3 &	0.23 & 0.05 & 16 &	0.25 & 0.04 & 7 &	0.08 & 0.04 & 8 &	0.29 & 0.04 & 11\\
 HD 224930   & 0.43 & 0.08 & 3 &	0.25 & 0.05 & 16 &	0.30 & 0.04 & 5 &	0.10 & 0.03 & 11&	0.27 & 0.04 & 9\\
\noalign{\smallskip}
\hline
 \noalign{\smallskip}
 HD 17548    & --0.02 & 0.01 & 2 &	0.08 & 0.05 & 17 &	0.10 & 0.04 & 7 &	0.05 & 0.04 & 10&	0.11 & 0.06 & 7\\
 HD 150177   & 0.06   & 0.04 & 2 &	0.05 & 0.06 & 12 &	0.10 & 0.03 & 5 &	0.08 & 0.03 & 7	&	0.15 & 0.05 & 3 \\
 HD 159307   & ...	  & ...  & ...&	0.16 & 0.02 &  9 &	0.17 & 0.04 & 5 &	0.12 & 0.06 & 7 &	0.17 & 0.07 & 3\\
 HD 165908   & 0.02   & 0.02 & 2 &	0.11 & 0.05 & 15 &	0.12 & 0.05 & 6 &	0.02 & 0.04 & 7 &	0.12 & 0.04 & 6\\
 HD 174912   & --0.03 & 0.04 & 2 &	0.04 & 0.04 & 17 &	0.09 & 0.05 & 6 &	0.00 & 0.04 & 12&	0.02 & 0.05 & 9\\
 HD 207978   & --0.01 & 0.02 & 2 &	0.12 & 0.04 & 15 &	0.15 & 0.03 & 7 &	0.06 & 0.02 & 6 &	0.16 & 0.06 & 3\\
\hline\hline   
Star & [Ti{\sc ii}/Fe] & $\sigma$ & n & [V/Fe] & $\sigma$ & n & [Cr/Fe] &$\sigma$ &  n& [Co/Fe]& $\sigma$& n& [Ni/Fe]& $\sigma$& n\\
 \hline
\noalign{\smallskip}
 HD 967      &   0.30 & 0.09 & 2 & 0.09 & 0.05 & 11 &	  0.05 & 0.08 & 17 &	0.08 & 0.04 & 5	&	0.02 & 0.07 & 26\\
 HD 17820    &   0.30 & 0.01 & 3 & 0.07 & 0.04 &  5 & 	--0.01 & 0.08 & 14 &	0.07 & 0.01 & 3 &	0.02 & 0.04 & 22\\
 HD 107582   & 	 0.27 & 0.06 & 2 & 0.05 & 0.03 & 10 &     0.05 & 0.06 & 14 &	0.04 & 0.05 & 9 &	0.02 & 0.06 & 16\\
 BD +73 566  &   0.35 & 0.06 & 3 &--0.03 & 0.04 &  6 &    0.05 & 0.05 & 11 &	0.06 & 0.04 & 3 &	0.00 & 0.05 & 17\\
 BD +19 2646 &   0.21 & 0.04 & 3 & 0.10 & 0.04 &  8 &	  0.06 & 0.10 & 15 &	0.05 & 0.06 & 7 &	0.00 & 0.05 & 18\\
 HD 114762   &   0.21 & 0.05 & 3 & 0.05 & 0.05 &  4 &	  0.00 & 0.08 & 11 &	0.04 & 0.05 & 5 & --0.04 & 0.04 & 16\\  
 HD 117858   &   0.20 & 0.03 & 3 & 0.11 & 0.03 &  8 &	  0.01 & 0.05 & 16 &	0.09 & 0.04 & 6 &	0.02 & 0.05 & 24\\
 BD +13 2698 &   0.33 & 0.05 & 3 & 0.08 & 0.03 &  7 &	  0.04 & 0.06 & 14 &	0.09 & 0.03 & 4 &	0.03 & 0.06 & 19\\
 BD +77 0521 &   0.17 & 0.03 & 2 & 0.06 & 0.02 &  3 &	  0.04 & 0.10 & 11 &	0.04 & 0.02 & 2 &	0.06 & 0.06 & 13\\
 HD 126512   &   0.29 & 0.02 & 2 & 0.08 & 0.03 &  7 &	  0.02 & 0.06 & 14 &	0.09 & 0.04 & 5 &	0.01 & 0.05 & 17\\
 HD 131597   &   0.30 & 0.03 & 3 & 0.06 & 0.06 & 11 &	  0.04 & 0.07 & 16 &	0.05 & 0.06 & 10&	0.01 & 0.05 & 25\\
 BD +67 925  &   0.43 & ...   & 1 &--0.03 & 0.08 &  4 &	--0.02 & 0.12 & 13 &	--0.02& 0.01 & 2 &	0.04 & 0.06 & 14\\
 HD 159482   &   0.23 & 0.01 & 3 & 0.09 & 0.03 &  7 &	  0.05 & 0.06 & 13 &	0.10 & 0.04 & 4 &	0.01 & 0.05 & 15\\
 HD 170737   &   0.28 & 0.06 & 4 & 0.08 & 0.05 &  8 &	  0.03 & 0.09 & 16 &	0.05 & 0.04 & 6 &	0.02 & 0.07 & 23\\
 BD +35 3659 &   0.24 & ...  & 1 &...   & ...  & ...&     0.03 & 0.11 & 12 &	...  & ...  &...&	0.01 & 0.08 & 7\\
 HD 201889   &   0.33 & 0.05 & 3 & 0.02 & 0.08 &  4 &     0.05 & 0.06 & 12 &	0.07 & 0.03 & 6 &	0.00 & 0.05 & 17\\
 HD 204521   &   0.29 & 0.04 & 3 & 0.06 & 0.05 &  7 &     0.02 & 0.06 & 13 &	0.07 & 0.05 & 6 &	0.00 & 0.05 & 18\\
 HD 204848   &   0.26 & 0.07 & 3 & 0.07 & 0.03 & 13 &	  0.06 & 0.09 & 17 &	0.05 & 0.03 & 10&	0.05 & 0.06 & 25\\
 HD 212029   &   0.36 & 0.03 & 2 & 0.09 & 0.05 & 4  &   --0.05 & 0.06 & 12 &	...  & ...  & ...&	0.01 & 0.06 & 11\\
 HD 222794   &   0.27 & 0.04 & 3 & 0.07 & 0.05 & 10 &	  0.03 & 0.07 & 16 &	0.06 & 0.03 & 8 &	0.02 & 0.06 & 18\\
 HD 224930   &   0.22 & 0.03 & 2 & 0.12 & 0.03 & 6  &   0.05 & 0.08 & 14 &	0.10 & 0.04 & 7 & --0.01 & 0.06 & 21\\
\noalign{\smallskip}
\hline
 \noalign{\smallskip}
 HD 17548    &   0.08 & 0.03 & 3 & 0.09 & 0.04 &  4 &	--0.01 & 0.05 & 13 &	0.05 & 0.03 & 3 & --0.04 & 0.05 & 20\\
 HD 150177   &   0.23 & 0.05 & 2 & 0.03 & ...  &  1 &   --0.03 & 0.07 & 14 &	0.06 & 0.00 & 2 & --0.02 & 0.06 & 12\\
 HD 159307   &   0.18 & 0.05 & 2 & ...  & ...  & ...&     0.04 & 0.05 &  9 &	0.06 & ...  & 1 &   0.04 & 0.02 & 10\\
 HD 165908   &   0.07 & 0.03 & 3 & 0.06 & 0.03 &  4 &	  0.00 & 0.04 & 10 &	0.00 & 0.08 & 5 & --0.03 & 0.06 & 15\\
 HD 174912   &   0.00 & 0.03 & 3 & --0.01 & 0.05 &  5 &	  0.00 & 0.08 & 14 &	0.04 & 0.00 & 4 & --0.05 & 0.05 & 21\\
 HD 207978   &   0.07 & 0.06 & 3 & ...    & ...  & ...&	  0.01 & 0.07 & 11 &	0.05 & 0.06 & 5 & --0.01 & 0.05 & 15\\
\hline
\end{tabular}
\end{minipage}
\end{table*}

The metallicities and ages of all programme stars except one (BD +35\,3659) are quite homogeneous: 
${\rm [Fe/H]}= -0.69\pm 0.05$~dex and the average age is about $12\pm 2$~Gyrs. However, the ages, which we took from 
\citet{holmberg09}, were not determined for every star.
BD +35 3659 is much younger (0.9~Gyr), has ${\rm [Fe/H]}= -1.45$, its eccentricity, velocities, distance, and other parameters 
differ as well (see Fig.~5 and 6). We doubt its membership of Group\ 3.
 
The next step was to compare the determined abundances with those in the thin-disc dwarfs. 
In Fig.~7 we present these comparisons with data taken from \citet{edvardsson93}, \citet{bensby05}, \citet{reddy06}, \citet{zhang06},
and with the chemical evolution model by \citet{pagel95}.
The thin-disc stars from \citeauthor{edvardsson93} and \citeauthor{zhang06} were selected by using the membership probability evaluation method described by \citet{trevisan11},
since their lists contained stars of other Galactic components as well. The same kinematical approach in assigning thin-disc membership was used in  
\citet{bensby05} and \citet{reddy06}, so the thin-disc stars used for the comparison are uniform in that respect. 
   
In Fig.~7 we  see that the abundances of $\alpha$-elements in the investigated stars are overabundant compared 
with the Galactic thin-disc dwarfs. A similar overabundance of $\alpha$-elements is exhibited by the thick-disc stars 
(\citealt{fuhrmann98, prochaska00, tautvaisiene01, bensby05, reddy08}; and references therein). \citet{helmi06}, based on the isochrone fitting, have suggested that stars in the 
identified kinematic groups might be $\alpha$-rich. Our spectroscopic results qualitatively agree 
with this. However, based on metallicities and vertical velocities, Group~3 cannot be uniquely 
associated to a single traditional Galactic component \citep{helmi06}.

What does the similarity of $\alpha$-element abundances in the thick-disc and the investigated kinematic group mean?
It would be easier to answer this question if the origin of the 
thick disc of the Galaxy was known (see \citealt{vanderkruit11} for a review). 

There are several competing models that aim to explain the nature of a thick disc. Stars may have appeared at the thick disc 
through
(1) orbital migration because of heating of a pre-existing thin disc by a varying gravitational potential in the thin disc 
(e.g. \citealt{roskar08, schonrich09}); 
(2) heating of a pre-existing thin disc by minor mergers (e.g. \citealt{kazantzidis08, villalobos08}, 2009);
(3) accretion of disrupted satellites (e.g. \citealt{abadi03}), or 
(4) gas-rich satellite mergers when thick-disc stars form before the gas completely settles into a thin-disc 
(see \citealt{brook04}, 2005). 

\citet{dierickx10} analysed the eccentricity distribution of thick-disc stars that has recently been proposed as a diagnostic to 
differentiate between these mechanisms \citep{sales09}. Using SDSS data release 7, they have assembled a sample 
of 31.535 G-dwarfs with six-dimensional phase-space information and metallicities and have derived their orbital eccentricities. 
They found that the observed eccentricity distribution is inconsistent with that predicted by orbital migration only. 
Also, the thick disc cannot be produced predominantly through heating of a pre-existing thin disc, since this model predicts more high-eccentricity stars than observed.
According to \citeauthor{dierickx10}, the observed eccentricity distribution  fits well with a gas-rich merger scenario, where most thick-disc stars were born in situ. 
In the gas-rich satellite merger scenario, a distribution of stellar eccentricities peak around $e=0.25$, with a tail towards higher 
values belonging mostly to stars originally formed in satellite galaxies. The group of stars investigated in our work fits this model 
with a mean eccentricity value of 0.4. This scenario is also supported by the RAVE survey data analysis made by 
\citet{wilson11} and the numerical simulations by \citet{dimatteo11}. In this scenario, Group~3 can be explained as 
a remnant from stars originally formed in a merging satellite.

\section{Conclusions}

We measured abundances of iron group and $\alpha$-elements from high-resolution spectra 
in 21 stars belonging to Group~3 of the Geneva-Copenhagen survey. This kinematically identified group of stars 
was suspected to be a remnant of a disrupted satellite galaxy.
Our main goal was to investigate the chemical composition of the stars within the group and to compare them with Galactic disc stars. 
         
Our study shows that 
  \begin{enumerate}
     \item All stars in Group~3 except one have a similar metallicity. The average 
     [Fe/H] value of the 20 stars is $-0.69\pm 0.05$~dex.
    \item  All programme stars are overabundant in oxygen and $\alpha$-elements compared with Galactic 
    thin-disc dwarfs and the Galactic evolution model used. This abundance pattern has similar characteristics as the 
    Galactic thick disc.  
  \end{enumerate} 

The homogeneous chemical composition together with the kinematic properties and ages of stars in the investigated Group~3 of 
the Geneva-Copenhagen
survey support the scenario of an ancient merging event. The similar chemical composition of stars in Group~3 
and the thick-disc stars might suggest that their formation histories are linked.
The kinematic properties of our stellar group fit well with a gas-rich satellite merger scenario (\citealt{brook04}, 2005; \citealt{dierickx10,
wilson11, dimatteo11}; and references therein).

We plan to increase the number of stars and chemical elements investigated in this group, and also to study the chemical composition of stars in other kinematic groups of the 
Geneva-Copenhagen survey. The identification of such kinematic groups and the exploration of their 
chemical composition will be a key in understanding the formation and evolution of the Galaxy. 

\begin{acknowledgements}
The data are based on observations made with the Nordic Optical Telescope, operated on the island of La Palma jointly by Denmark, Finland, Iceland,
Norway, and Sweden, in the Spanish Observatorio del Roque de los Muchachos of the Instituto de Astrofisica de Canarias.  
The research leading to these results has received funding from the European Community's Seventh Framework Programme (FP7/2007-2013) under
grant agreement number RG226604 (OPTICON). BN acknowledges support from the Danish Research council. 
This research has made use of Simbad, VALD and NASA ADS databases. We thank the anonymous referee for 
insightful questions and comments.
\end{acknowledgements}

\end{document}